\documentclass[%
	final,
	conference, 
    comsoc,
	letterpaper,
	oneside,
	twocolumn,
	nofonttune,%
]{IEEEtran}%
\PassOptionsToPackage{usenames,dvipsnames,svgnames,x11names,table,prologue}{xcolor}%
\PassOptionsToPackage{hyphens}{url}%
\PassOptionsToPackage{nomessages}{fp}%
\ifCLASSINFOpdf
  \usepackage[pdftex]{graphicx}
\else
  \usepackage[dvips]{graphicx}
\fi
\ifCLASSOPTIONcompsoc
   \usepackage[caption=false,font=normalsize,labelfont=sf,textfont=sf]{subfig}
\else
   \usepackage[caption=false,font=footnotesize]{subfig}
\fi

\usepackage{stfloats}
\usepackage[english]{babel}%
\usepackage[utf8]{inputenc}%
\usepackage[babel,style=english]{csquotes}%
\usepackage{hyphsubst}%
\usepackage[%
	activate={true,%
	nocompatibility},%
	final,%
	tracking=true,%
	kerning=true,%
	spacing=true,%
	factor=1100,%
	stretch=10,%
	shrink=10%
]{microtype}%
\usepackage{setspace}%
\usepackage{xcolor}%
\definecolor{todonotecol}{RGB}{250,0,0}%
\usepackage{xparse}
\usepackage[%
	colorlinks=false,%
	urlcolor=black,%
	linkcolor=black,%
	citecolor=black,%
	filecolor=black,%
	breaklinks,%
	]{hyperref}%
\usepackage{url}%
\usepackage[%
	acronym,%
	nopostdot,%
	seeautonumberlist,%
	shortcuts,%
	section=chapter,%
	toc,%
]{glossaries}%
\glsaddkey*{url}
{}
{\glsentryurl}
{\Glsentryurl}
{\glsurl}
{\Glsurl}
{\GLSurl}
\glsaddkey*{longpltwo}
{}
{\glsentrylongpltwo}
{\Glsentrylongpltwo}
{\glslongpltwo}
{\Glslongpltwo}
{\GLSlongpltwo}
\newglossaryentry{}{%
	name={},%
	plural={},%
	description={},%
}%
\newacronym{it}{IT}{information technology}
\newacronym{ot}{OT}{operational technology}
\newacronym{opcua}{OPC UA}{Open Platform Communications Unified Architecture}
\newacronym{tsn}{TSN}{Time-Sensitive Networking}
\newacronym{pc}{PC}{Personal Computer}
\newacronym{ptp}{PTP}{Precision Time Protocol}
\newacronym{gptp}{gPTP}{generalized Precision Time Protocol}
\newacronym{ntp}{NTP}{Network Time Protocol}
\newacronym{tacnet4.0}{TACNET 4.0}{Tactile Internet 4.0}
\newacronym{3gpp}{3GPP}{3rd Generation Partnership Project}
\newacronym{5gppp}{5GPPP}{5G Infrastructure Public Private Partnership}
\newacronym{itu}{ITU}{International Telecommunication Union}
\newacronym{ngnm}{NGNM}{Next Generation Mobile Networks}
\newacronym{5g}{5G}{Fifth Generation Mobile Network}
\newacronym{harq}{HARQ}{Hybrid Automatic Repeat Request}
\newacronym{rat}{RAT}{radio access technology}
\newacronym{rt}{RT}{Real-Time}
\newacronym{wlan}{WLAN}{Wireless Local Area Network}
\newacronym{cpu}{CPU}{central processing unit}
\newacronym{ram}{RAM}{random-access memory}
\newacronym{ue}{UE}{User Equipment}
\newacronym{ran}{RAN}{Radio Access Network}
\newacronym{bs}{BS}{base station}
\newacronym{cn}{CN}{core network}
\newacronym{epc}{EPC}{evolved packet core}
\newacronym{mec}{MEC}{mobile edge cloud}
\newacronym{rtt}{RTT}{Round Trip Time}
\newacronym{iot}{IoT}{Internet of Things}
\newacronym{urllc}{URLLC}{ultra-reliable low-latency communication}
\newacronym{mmtc}{mMTC}{massive machine-type communications}
\newacronym{iiot}{IIoT}{Industrial Internet of Things}
\newacronym{npn}{NPN}{non-public network}
\newacronym{prb}{PRB}{Physical Ressource Block}
\newacronym{gnb}{gNB}{5G Base Station}
\newacronym{5gs}{5GS}{5G System}
\newacronym{5gc}{5GC}{5G Core Network}
\newacronym{oai}{OAI}{OpenAirInterface}
\newacronym{sdr}{SDR}{Software Defined Radio}
\newacronym{usrp}{USRP}{Universal Software Radio Peripheral}
\newacronym{mimo}{MIMO}{Multiple Input Multiple Output}
\newacronym{fpga}{FPGA}{Field Programmable Gate Array}
\newacronym{pss}{PSS}{Primary Synchronization Signal}
\newacronym{sss}{SSS}{Secondary Synchronization Signal}
\newacronym{sfn}{SFN}{System Frame Number}
\newacronym{ssb}{SSB}{Synchronization Signal Block}
\newacronym{mib}{MIB}{Master Information Block}
\newacronym{pbch}{PBCH}{Physical Broadcast Channel}
\newacronym{ism}{ISM}{Industrial, Scientific and Medical}
\newacronym{dl}{DL}{Downlink}
\newacronym{ds-tt}{DS-TT}{Device-Side TSN Translator}
\newacronym{nw-tt}{NW-TT}{Network-Side TSN Translator}
\newacronym{upf}{UPF}{User Plane Function}
\newacronym{tsi}{TSi}{ingress timestamping}
\newacronym{tse}{TSe}{egress timestamping}
\newacronym{gm}{GM}{grandmaster}
\newacronym{sn}{SN}{sequence number}
\newacronym{rbis}{RBIS}{Reference Broadcast Infrastructure Synchronization}
\newacronym{cots}{COTS}{Commercial Off-The-Shelf}
\newacronym{b5g}{B5G}{Beyond 5G}
\newacronym{msc}{MSC}{Message Sequence Chart}
\newacronym{ta}{TA}{Timing Advance}
\newacronym{tdd}{TDD}{Time Division Duplex}
\glsdisablehyper
\usepackage[%
	backend=biber,%
	style=ieee,%
	isbn=false,%
	hyperref=true,%
	maxbibnames=99,%
	sorting=none,%
	natbib=true,%
	language=english,%
	defernumbers=true,%
	]{biblatex}%
\DeclareFieldFormat{sentencecase}{\csname bbx@colon@search\endcsname#1}

\usepackage{nameref}
\usepackage{soul}
\usepackage{flushend}
\usepackage{textcomp}
\usepackage{calc}
\usepackage{xkeyval}
\usepackage{multirow}
\usepackage{tabulary}
\usepackage{makecell}
\usepackage{pgfplots}
\usepackage{tikz}
\usepackage{textcomp} 
\usepackage{verbatim}
\newcommand{\mytilde}{{\raise.17ex\hbox{$\scriptstyle\mathtt{\sim}$}}}

\newlength\textheighttemp%
\newlength\textwidthtemp%
\newlength\textheightstd%
\newlength\textwidthstd%
\newlength\textheightold%
\newlength\textwidthold%
\newlength\tempheight%
\newlength\tempwidth%
\SetProtrusion{encoding={*},family={bch},series={*},size={6,7}}
              {1={ ,750},2={ ,500},3={ ,500},4={ ,500},5={ ,500},
               6={ ,500},7={ ,600},8={ ,500},9={ ,500},0={ ,500}}
\SetExtraKerning[unit=space]
    {encoding={*}, family={bch}, series={*}, size={footnotesize,small,normalsize}}
    {\textendash={400,400}, 
     "28={ ,150}, 
     "29={150, }, 
     \textquotedblleft={ ,150}, 
     \textquotedblright={150, }} 
\SetTracking{encoding={*}, shape=sc}{40}
\makeatletter
\let\blx@rerun@biber\relax
\makeatother

\makeatletter
    \let\thanks\@IEEESAVECMDthanks%
\makeatother
\usepackage{listings}
\usepackage[dvipsnames]{xcolor}
\usepackage{tikz} \usetikzlibrary{calc, arrows.meta, intersections, patterns, positioning, shapes.misc, fadings, through,decorations.pathreplacing}

\usepackage{algorithm}
\usepackage{algpseudocode}
\usepackage{tabularx}
\usepackage{newtxmath}
\usepackage[percent]{overpic} 
\usepackage{booktabs}
\usepackage{balance}
\pgfplotsset{
  grid style = {
   line width = 0.1pt
  }
}
\pgfplotsset{compat=1.18}
				\newcommand{\disablewr}[1]{#1}%
				\newcommand{\newcommanddisw}[3]{\newcommand{#1}[1]{\disablewr{\textcolor{#2}{#3}}}}%
\renewcommand{\disablewr}[1]{}%
\definecolor{todocol}{named}{red}
\newcommanddisw{\todo}{todocol}{ToDo: #1}%
\definecolor{migucol}{named}{purple}%
\newcommanddisw{\migucom}{migucol}{{@}comment: #1}%
\newcommanddisw{\miguhigh}{migucol}{#1}%
\usepackage{siunitx}
\usepackage{xspace}

\begin{document}%
%
\title{%
Cracking the Microsecond: An Ef{}f{}icient and Precise Time Synchronization Scheme for Hybrid 5G-TSN Networks\thanks{This research was supported by the German Federal Ministry of Research, Technology and Space (BMFTR) within the projects Open6GHub and 6GTerafactory under grant numbers 16KISK003K and 16KISK186. The responsibility for this publication lies with the authors. This is a preprint of a work accepted but not yet published at the 2025 IEEE International Symposium on Precision Clock Synchronization for Measurement, Control, and Communication (ISPCS). Please cite as: M. Gundall and H. D. Schotten: “Cracking the Microsecond: An Ef{}f{}icient and Precise Time Synchronization Scheme for Hybrid 5G-TSN Networks”. In: 2025 IEEE International Symposium on Precision Clock Synchronization for Measurement, Control, and Communication (ISPCS), IEEE, 2025.}
}
%
\author{%
\IEEEauthorblockN{%
    Michael Gundall\IEEEauthorrefmark{1} %
    and Hans D. Schotten\IEEEauthorrefmark{1}\IEEEauthorrefmark{3} %
    \\%
}%
\IEEEauthorblockA{%
    \IEEEauthorrefmark{1}German Research Center for Artificial Intelligence (DFKI), Kaiserslautern, Germany \\%
    \IEEEauthorrefmark{3}Department of Electrical and Computer Engineering, RPTU University Kaiserslautern-Landau, Kaiserslautern, Germany %
	\\%
    Email: %
        \{michael.gundall, hans\_dieter.schotten%
        \}@dfki.de
        \\%
}%
}%

%
%
%
%
%
%
%
%
\maketitle
%
%
%
%
%
\begin{abstract}%
Achieving precise time synchronization in wireless systems is essential for both industrial applications and \acrshort{5g}, where sub-microsecond accuracy is required. 
However, since the \gls{iiot} market is negligible compared to the consumer electronics market, the so-called \gls{iiot} enhancements have not yet been implemented in silicon. Moreover, there is no guarantee that this situation will change soon. Thus, alternative solutions must be explored.

This paper addresses this challenge by introducing a scheme that uses a protocol capable of leveraging existing infrastructure to synchronize \glspl{ue}, with one of the \glspl{ue} serving as the master. If this master is connected via a wired link to the factory network, it can also function as a boundary clock for the factory network, including any \gls{tsn} network. Furthermore, the \gls{5gc} and \gls{gnb} can also be synchronized if they are connected either to the factory network or to the master \gls{ue}.

The proposed solution is implemented and evaluated on a hardware testbed using \gls{oai} and \glspl{sdr}. Time offset and clock skew are analyzed using a moving average filter with various window sizes. Results show that a filter size of 1024 provides the best accuracy for offset prediction between \glspl{ue}. In a controlled lab environment, the approach consistently achieves synchronization within ±50~ns, leaving sufficient margin for synchronization errors in real deployments while still maintaining sub-microsecond accuracy. These findings demonstrate the feasibility and high performance of the proposed protocol for stringent industrial use cases.
\end{abstract}%
%
%
%
%
%
\IEEEpeerreviewmaketitle
%
%
%
%
%
%
%
%
\tikzstyle{descript} = [text = black,align=center, minimum height=1.8cm, align=center, outer sep=0pt,font = \footnotesize]
\tikzstyle{activity} =[align=center,outer sep=1pt]

\section{Introduction}%
\label{sec:Introduction}
\glsresetall
Precise time synchronization is a fundamental requirement of modern information and communication technology (ICT) systems on earth and beyond~\cite{7772527}. This is especially true for industrial applications, where stringent \gls{rt} performance is crucial. In particular, synchronicity of less than \SI{1}{\micro\second} is mandated for demanding \gls{rt} class C use cases, ensuring deterministic behavior in automated industrial environments~\cite{7883994}.For emerging \acrshort{5g}-Advanced and 6G use cases, the requirement of one microsecond remains relevant~\cite{chandramouli2023evolution}. Additionally, precise synchronization is necessary to avoid interference in \gls{tdd} systems, which are increasingly used, for example, in frequency band n78, which serves for private networks in Germany, and in FR2 (mmWave) bands.

While wireline solutions, such as the White Rabbit extension of the \gls{ptp}, can achieve synchronization accuracies down to \SI{100}{\pico\second}~\cite{6070148}, extending this level of precision to wireless systems remains challenging. The inherent characteristics of wireless communication, such as mobility, varying channel conditions, packet loss and line-of-sight constraints, present challenges in meeting the requirements~\cite{chen2021understanding}.

In response to these challenges, \acrshort{5g} Release 16 and 17 specifications introduce concepts aimed at fulfilling the synchronization requirements for \gls{iiot}. However, practical implementation by chipset manufacturers is unlikely, as the \gls{iiot} market segment often lacks the scale and profit margins necessary to justify the investment, especially compared to the consumer market. Consequently, alternative methods must be explored to meet stringent timing demands of mobile use cases.

One potential approach is the tunneling of \gls{ptp} messages within \acrshort{5g} data packets~\cite{10710776}. Nevertheless, this method typically relies on software timestamps, in order to use existing hardware. Consequently, only millisecond-level accuracy is reached using this approach, falling short of industrial synchronization requirements~\cite{10710776}. Another promising alternative leverages the broadcast nature of wireless systems, which, while often considered a drawback due to interference and lack of determinism, can be exploited effectively for one-way time synchronization protocols. In this context, the so-called \gls{rbis} protocol is a viable solution~\cite{7018946}. Originally developed for Wi-Fi networks, \gls{rbis} is adaptable to any infrastructure-based wireless communication system, including \acrshort{5g}. Moreover, \gls{rbis} requires no modifications to existing network infrastructure, allowing seamless integration with standard \glspl{gnb}.

Additionally, \gls{rbis} supports the inclusion of external time sources, facilitating the realization of hybrid networks that converge \acrshort{5g} and  \gls{tsn}, as found in modern smart factories. Owing to these strengths, \gls{rbis} serves as the foundation for the investigations presented in this paper, which aims to evaluate its feasibility and performance in meeting the stringent synchronization requirements of industrial \acrshort{5g} use cases.

Accordingly, the following contributions can be found in this paper:
\begin{itemize}
  \item Scheme for efficient and precise time synchronization in hybrid \acrshort{5g}-\gls{tsn} networks.
  \item Evaluation on a testbed demonstrating sub-microsecond accuracy and precision.
\end{itemize}

Accordingly, the rest of the paper is structured as follows: Section~\ref{sec:Related Work} introduces related work. Furthermore, the proposed Scheme for achieving highly synchronized \acrshort{5g}-\gls{tsn} networks is presented in Section~\ref{sec:rbis}. Then, the experimental setup is introduced in Section~\ref{sec:Methodology}. Thereafter, the used methodology is explained in in Section~\ref{sec:Experimental Setup} and the results are discussed in Section~\ref{sec:Evaluation}. Finally, Section~\ref{sec:Conclusion} concludes the paper.

\section{Related Work}%
\label{sec:Related Work}
Time synchronization in cellular systems gained attention with Industry 4.0 use cases requiring sub-microsecond accuracy for high-performance wireless communications~\cite{etfa2018, mahmood2019time}. 3GPP Release 16 introduced \acrshort{5g} synchronization targets of \SI{1}{\micro\second}~\cite{godor2020look}. Both \gls{ds-tt} and \gls{nw-tt} have been proposed to integrate \acrshort{5g} in factory networks~\cite{nokiaWhitPaper}.

Research on applying \gls{ptp} in \acrshort{5g} revealed millisecond-level precision limitations due to tunneling in the data plane~\cite{10710776}. Other studies reported synchronization accuracy between \gls{ue} and \gls{gnb} in the \SI{470}-\SI{540}{\nano\second} range~\cite{9652097}, though lacking testbed evaluations. The authors in~\cite{kehl2022prototype} implemented a \gls{tsn} translator hardware testbed achieving \SI{1}-\SI{10}{\micro\second} accuracy. Propagation delay compensation as well as elimination of other error sources remains challenging, as discussed in~\cite{striffler2021}.

Most industrial devices are serving as \glspl{ue}, making synchronization between them more important than of a \gls{ue} with the \gls{gnb} from an application point of view. Thus, the \gls{rbis} protocol, originally proposed for Wi-Fi, is well suited for this purpose~\cite{7018946}. Prior work achieved sub-millisecond requirements~\cite{gundall2020integration} and $\pm$\SI{10}{\micro\second} accuracy with offset correction in distributed setups~\cite{10008489}. To our knowledge, consistent microsecond-level synchronization has not yet been demonstrated. Our work attempts to improve on this by applying both offset and rate correction, leveraging precise hardware timestamps from  \gls{sdr} clocks, which are more stable than CPU clocks that were used before.

\section{Efficient and Precise Time Synchronization Scheme for Hybrid \acrshort{5g}-\gls{tsn} Networks} %
\label{sec:rbis}
This section introduces the concept that enables efficient and precise time synchronization in hybrid \acrshort{5g}–\gls{tsn} networks. First, the \gls{rbis} protocol, which serves as the basis for this approach, is presented in Section~\ref{subsec:RBIS protocol}. Although it has certain prerequisites, it is technology independent and its mapping to \acrshort{5g} systems is described in Section~\ref{subsec:Mapping to 5G}. Furthermore, protocol extensions that make it well suited for hybrid \acrshort{5g}–\gls{tsn} networks as well as the architectural integration possibilities are proposed in Section~\ref{subsec:Extension to hybrid Networks}.

\subsection{\gls{rbis} Protocol}
\label{subsec:RBIS protocol}

\begin{figure}[htbp]
\centering
 \includegraphics[width=0.999\columnwidth]{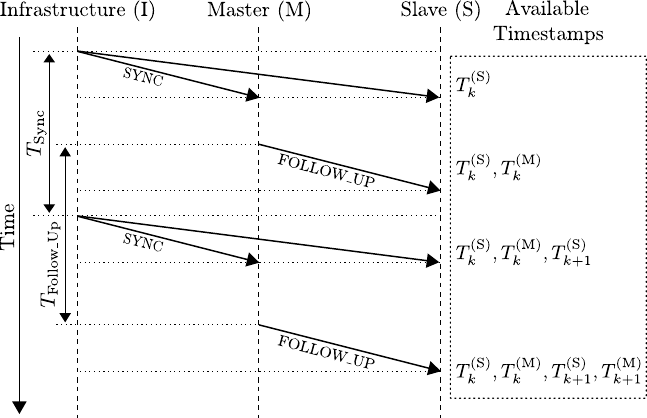}
\caption{\gls{msc} of technology-independent \gls{rbis} protocol.}
\label{fig:rbis_msc}
\end{figure}

Figure~\ref{fig:rbis_msc} depicts the \gls{msc} of the protocol.
Upon reception of a SYNC signal transmitted by the infrastructure, each device generates a timestamp based on its individual clock. Subsequently, the master transmits its \grqq correct" timestamps, denoted by $T_k^\mathrm{(M)}$, to all slaves via the FOLLOW\_UP message. Subsequently, each slave generates its own reference timestamps, denoted by $T_k^\mathrm{(S)}$. The resulting $T_k^\mathrm{(M)}$ and $T_k^\mathrm{(S)}$ tuples are used to estimate the offset $\theta$ and skew $\gamma$ between the master and slave clocks following Equations~\mbox{\ref{eq:1}-\ref{eq:2}}~\cite{6817598}.
\begin{equation}
\hat{\theta}_K=T_k^\mathrm{(S)}-{T}_k^\mathrm{(M)}
\label{eq:1}
\end{equation}
\begin{equation}
\hat{\gamma}_k=\frac{\hat{\theta}_k-\hat{\theta}_{k-1}}{T_k^\mathrm{(M)}-T_{k-1}^\mathrm{(M)}}
\label{eq:2}
\end{equation}
Additionally, the repetition intervals of both SYNC and FOLLOW\_UP messages do not have to be the same.

\subsection{Mapping to \acrshort{5g}}
\label{subsec:Mapping to 5G}
In order to make the \gls{rbis} protocol applicable to \acrshort{5g}, a potential synchronization signal has to be identified that serves as SYNC message and must contain a unique identifier that allows to differ upcoming from preceding signals. Here, the so-called \gls{pbch}, which is co-located to \gls{pss} and \gls{sss}, and can be found inside the \gls{ssb}, is well suited. Moreover, the \gls{pbch} carries the so-called \gls{sfn}. The \gls{sfn} is of interest because it is a cyclically incremented identifier. Furthermore, it is incremented every  \SI{1}{\milli\second} and has a length of  \SI{10}{\bit}. So the maximum value is $1023$ and a repetition occurs every \SI{10.24}{\second}. This means that the second criterion for assigning a timestamp to an unique message can only be partially fulfilled, however, an initial offset of $\theta_\mathrm{0} < 10.24/2~\mathrm{s}  =$  \SI{5.12}{\second} can easily guaranteed by sending an initial timestamp $\theta_\mathrm{0}$.
Moreover, the \gls{pbch} is sent in an interval  between  \SI{5}{\milli\second} and  \SI{160}{\milli\second} and can serve as $T_\mathrm{Sync}$ variable of the \gls{rbis} protocol. 

\subsection{Protocol Extensions and Architectural Integration}
\label{subsec:Extension to hybrid Networks}
\begin{figure}[htbp]
\centering
 \includegraphics[scale=0.9]{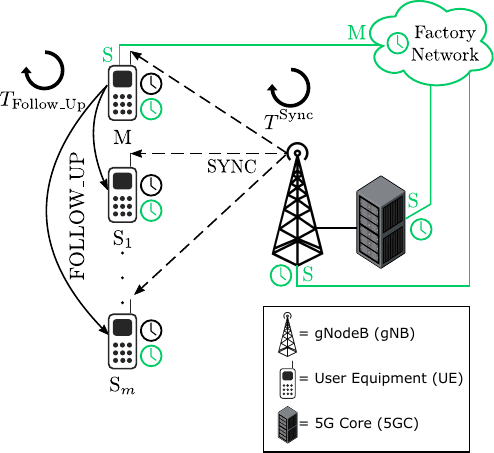}
\caption{Architectural integration possibilities of the \gls{rbis} protocol in hybrid \acrshort{5g}-\gls{tsn} networks.}
\label{fig:rbis}
\end{figure}

This section introduces the existing \acrshort{5g}-specific protocol extensions and suggests how the \gls{rbis} protocol can be integrated into hybrid networks. As illustrated in Figure~\ref{fig:rbis}, 
 the \gls{rbis} protocol does not compensate the signal propagation time, which is negligible in Wi-Fi deployments, due to the typically small spatial coverage. However, in \acrshort{5g} scenarios where larger spatial coverage is feasible, variations in the distance between the \glspl{ue} and the \gls{gnb} can introduce runtime errors in the synchronization. To mitigate this, the so-called \gls{ta}, which is calculated by the \gls{gnb} and transmitted to the \glspl{ue}, can be utilized to compensate for propagation-induced delays and improve synchronization accuracy~\cite{mahmood2019time}. 

Furthermore, in its current form, the protocol synchronizes only the \glspl{ue} within a single cell. To extend synchronization coverage beyond a single cell, concepts for expanding the coverage area both non-invasively and invasively have been proposed, with the latter approach aligning with the evolution towards future 6G networks~\cite{10217973}.

In order to synchronize the \glspl{ue} to a \gls{tsn} master clock, the master \gls{ue} can be connected wireline to the factory network. Hence, a single \gls{ue} needs to be stationary, but can serve as boundary clock between \glspl{ue} and \gls{tsn} components. Consequently, the time of a \gls{tsn} master located in the factory network can also be distributed to all \glspl{ue}. Thus, the remaining limitation of the protocol, namely its inability to synchronize \glspl{ue} with \gls{gnb} and \gls{5gc}, is lifted.

\section{Experimental Setup}%
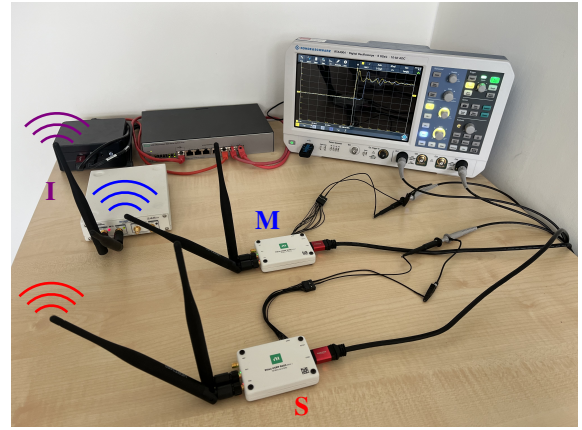
\begin{figure}[htbp]
\centering
 \begin{overpic}[width=0.85\columnwidth]{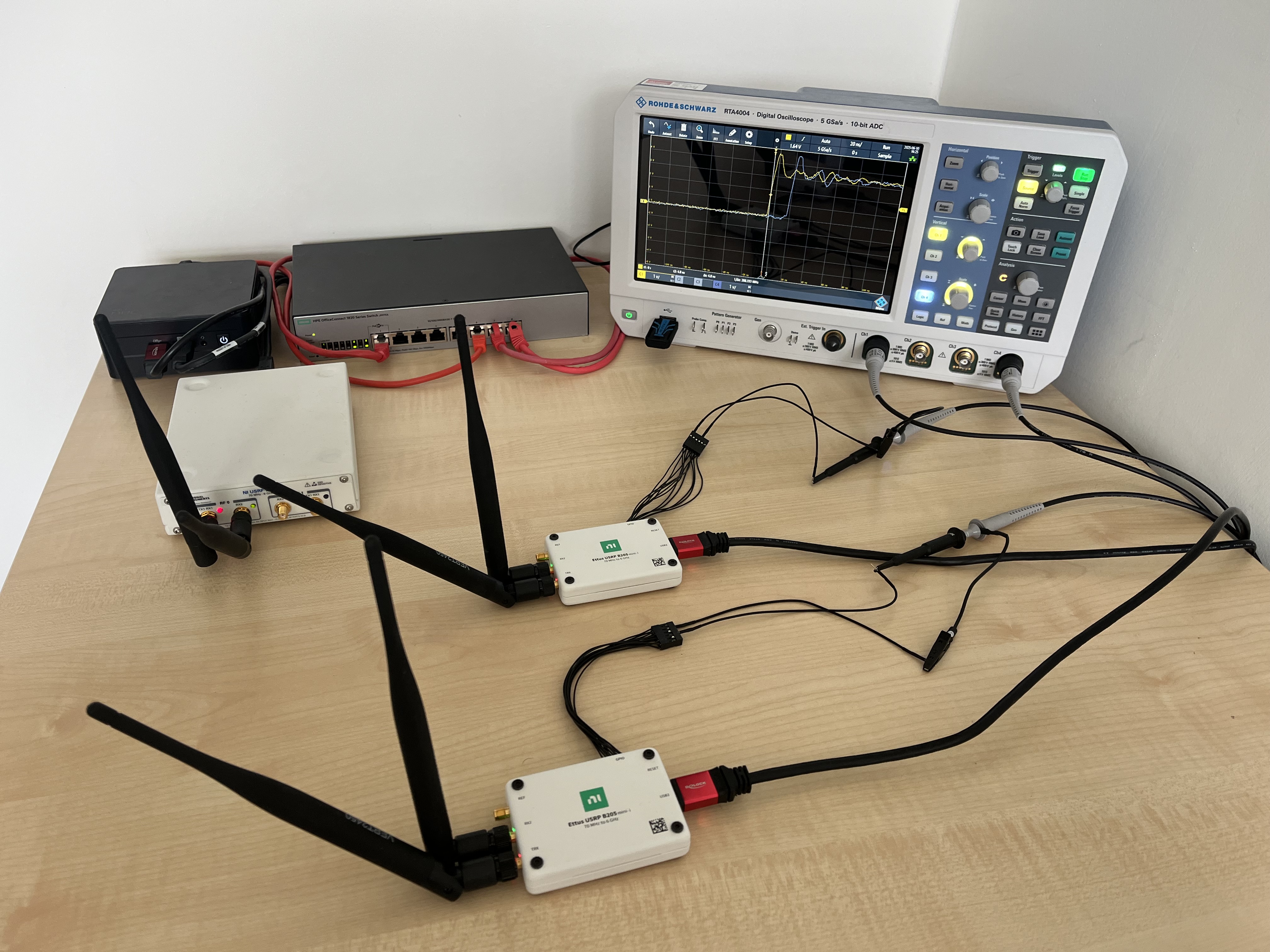}
  \put(50,2){\color{red}\textbf{S}}
    \put(0,20){ 
      \begin{tikzpicture}[scale=0.3]
        \draw[red,line width=1pt] (135:2) arc[start angle=135,end angle=45,radius=2];
        \draw[red,line width=1pt] (135:1.5) arc[start angle=135,end angle=45,radius=1.5];
        \draw[red,line width=1pt] (135:1) arc[start angle=135,end angle=45,radius=1];
      \end{tikzpicture}
      }
        \put(6,40){\color{violet} \textbf{I}}
    \put(1,50){ 
      \begin{tikzpicture}[scale=0.3]
        \draw[violet, line width=1pt] (135:2) arc[start angle=135,end angle=45,radius=2];
        \draw[violet, line width=1pt] (135:1.5) arc[start angle=135,end angle=45,radius=1.5];
        \draw[violet, line width=1pt] (135:1) arc[start angle=135,end angle=45,radius=1];
      \end{tikzpicture}
      }
        \put(43,35){\color{blue} \textbf{M}}
    \put(13,38){ 
      \begin{tikzpicture}[scale=0.3]
        \draw[blue, line width=1pt] (135:2) arc[start angle=135,end angle=45,radius=2];
        \draw[blue, line width=1pt] (135:1.5) arc[start angle=135,end angle=45,radius=1.5];
        \draw[blue, line width=1pt] (135:1) arc[start angle=135,end angle=45,radius=1];
      \end{tikzpicture}
      }
 \end{overpic}
\caption{Experimental setup consisting of a mini \gls{pc}, one \gls{usrp} B210 as \gls{gnb}, two \gls{usrp} B205mini as \glspl{ue}, and an oscilloscope.}
\label{fig:OverviewSetup}
\end{figure}

\label{sec:Experimental Setup}
The experimental setup is shown in Figure~\ref{fig:OverviewSetup}.
It consists of general-purpose computing hardware,  \glspl{sdr}, and measurement equipment, as summarized in Table~\ref{tab:hardware}.
\begin{table}[tb]
    \centering
    \caption{Hardware configurations}
    \begin{tabular}{c   c   l }
    \toprule
\textbf{Equipment} & \textbf{QTY} & \textbf{Specification}\\
     \midrule
Mini PC & 1 & 13\textsuperscript{th} Gen. Intel Core i7-1360P, \\
& & 32 GB DDR4, 3x USB 3.2 Gen2 \\
& & Ubuntu 24.04.2 LTS 64-bit \\
5G gNB & 1 & Ettus Research USRP B210 \\
5G UE & 2 & Ettus Research USRP B205mini \\
\midrule
Oscilloscope & 1 & 	Rohde\&Schwarz 	RTA4004\\
      \bottomrule 
    \end{tabular}
    \label{tab:hardware}
\end{table}
The core processing unit is a mini \gls{pc} equipped with an Intel Core i7 CPU and 32 GB of DDR4 RAM. The system runs Ubuntu~24.04.2~LTS~(64-bit). The entire 5G sysem is implemented using the \gls{oai}\footnote{https://gitlab.eurecom.fr/oai/openairinterface5g/} open-source software suite. Specifically, the \gls{5gc}, \gls{gnb}, and \gls{ue} functionalities are all executed concurrently on the same mini \gls{pc}, enabling an integrated and reproducible \acrshort{5g} setup. This is possible, as the selected mini \gls{pc} has three high speed USB ports.

For the radio access network, an Ettus Research \gls{usrp} B210  \gls{sdr} is used as \gls{gnb}, denoted by I. It supports RF frequencies from  \SI{70}{\mega\hertz} to  \SI{6}{\giga\hertz} with a 2×2 MIMO configuration and a Spartan 6 FPGA. Hence, they are suitable to operate in frequency band n78 that is used for private networks in Germany. Moreover, two Ettus Research \gls{usrp} B205mini  \glspl{sdr} serve as \glspl{ue} (M and S), which provide similar RF capabilities and, importantly, feature accessible GPIO pins. One GPIO pin from each B205mini device is connected to a channel of a Rohde~\&~Schwarz~RTA4004 oscilloscope, which operates with a maximum sampling rate of 
$5$~GSa/s. This enables high-resolution measurement of the generated validation signals, allowing precise offset measurements to evaluate the time synchronization error between both \glspl{ue}.

\section{Methodology}%
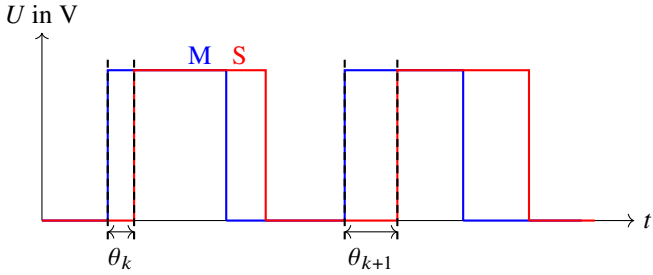
\begin{figure}
\centering
    \begin{tikzpicture}[xscale=3.5,yscale=1]
    \draw[->](-0.25,0)--(2,0)node[right]{$t$};
    \draw[->](-0.25,0)--(-0.25,2.5)node[above]{$U$ in V};
    \def\T{0.9}
    \def\D{0.5}
    \def\Ttwo{1.0}
    \def\Dtwo{0.5}
    \def\shift{0.1}
    \def\N{1}
    \foreach \i in {0,...,\N}{
      \draw[thick,blue]({\i*\T},0)--({\i*\T},2)--({\i*\T+\D*\T},2)--({\i*\T+\D*\T},0);
      \draw[thick, blue]({\i*\T+\D*\T},0)--({(\i+0.5)*\T+\D*\T},0);
    }
    \node[blue] at (0.5*\T-\shift,2.2) {M};
    \draw[thick, blue](-0.25,0)--(0,0);
    \foreach \i in {0,...,\N}{
      \draw[thick,red]({\i*\Ttwo+\shift},0)--({\i*\Ttwo+\shift},2)--({\i*\Ttwo+\Dtwo*\Ttwo+\shift},2)--({\i*\Ttwo+\Dtwo*\Ttwo+\shift},0);
    }
    \draw[thick,red]({0*\Ttwo+\Dtwo*\Ttwo+\shift},0)--({(0.5)*\Ttwo+\Dtwo*\Ttwo+\shift},0);
    \draw[thick,red]({1*\Ttwo+\Dtwo*\Ttwo+\shift},0)--({(1.25)*\Ttwo+\Dtwo*\Ttwo+\shift},0);
    \node[red] at (0.5*\Ttwo,2.2) {S};
    \draw[thick,red](-0.25,0)--(\shift,0);
    \draw[<->](0,-0.15)--(\shift,-0.15);
    \node at (0.5*\shift,-0.5){$\theta_k$};
    \draw[<->](1-\shift,-0.15)--(1+\shift,-0.15);
    \node at (1,-0.5){$\theta_{k+1}$};
    \draw[thick,black,densely dashed](0,-0.25)--(0,2.2);
    \draw[thick,black,densely dashed](\shift,-0.25)--(\shift,2.2);
    \draw[thick,black,densely dashed](1-\shift,-0.25)--(1-\shift,2.2);
    \draw[thick,black,densely dashed](1+\shift,-0.25)--(1+\shift,2.2);
    \end{tikzpicture}
\caption{Offset between two rectangular signals over time.}
\label{fig:validation}   
\end{figure}

\label{sec:Methodology}
To evaluate the clock synchronization performance between two \glspl{ue}, a measurement procedure is implemented to estimate clock skew and clock offset over time. To mitigate short-term noise and measurement fluctuations, a moving average filter of size $N$ is applied to estimate the underlying clock skew. Using this prediction, the offset at a future time instant can then be determined by integrating the estimated drift over time and adding it to the last measured offset. To evaluate the achievable accuracy and precision, each \gls{ue} periodically generates a rectangular signal via its accessible GPIO pin. As shown in Figure~\ref{fig:validation},
two such signals, one from each \gls{ue}, are generated with voltage levels varying between  \SI{0}{\volt} (logical 0) and  \SI{3.3}{\volt} (logical 1). These signals are generated on the \glspl{usrp} using timed commands to ensure deterministic signal generation aligned with their hardware clock. Both signals are captured simultaneously using the oscilloscope. By comparing the time stamps of rising or falling edges of the two validation signals, the instantaneous clock offset between the \glspl{ue} can be determined. Repeated measurements over an extended period allow observation of how this offset evolves over time due to clock skew.

\section{Evaluation}%
\label{sec:Evaluation}
In this section, the proposed time synchronization scheme is evaluated in detail. First, the clock skew is identified in Section~\ref{subsec:skew prediction}, and based on these measurements, an appropriate filter size is selected and integrated into the algorithm. This is followed by the presentation of offset measurement results in Section~\ref{subsec:results}, which demonstrate precision and accuracy of the synchronization scheme.

\subsection{Identification of Clock Skew}
\label{subsec:skew prediction}
\begin{figure}[htbp]
	\centering
		\subfloat[Time series obtained for the readings of estimated clock skew $\hat{\gamma}$.]{\resizebox{0.99\columnwidth}{!}{\input{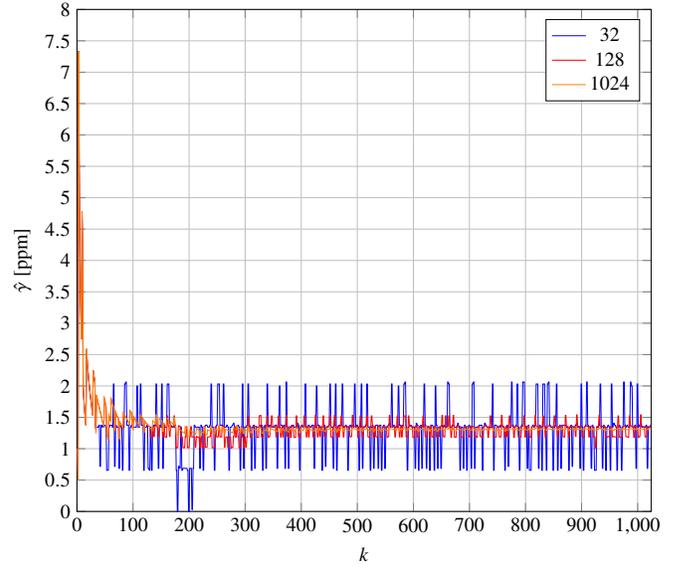}}\label{fig:1_1}}
		
         \subfloat[Distribution functions obtained for the readings of estimated clock skew $\hat{\gamma}$.]{\resizebox{0.99\columnwidth}{!}{\input{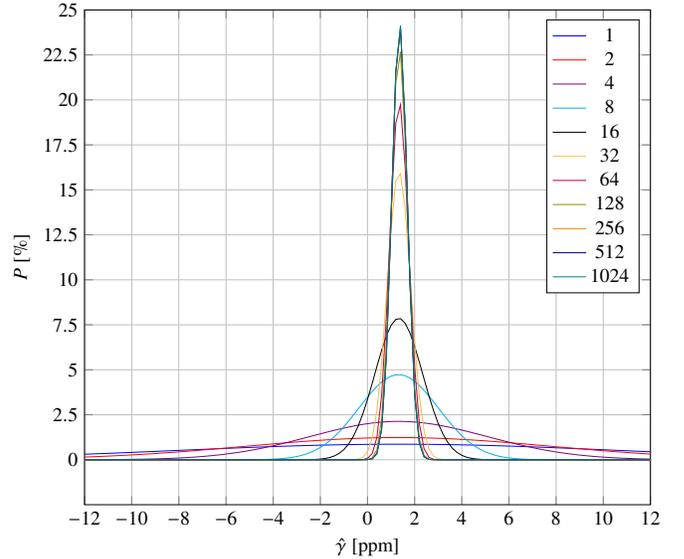}}\label{fig:1_2}}
\caption{Illustration of the convergence behavior of the clock skew estimation with increasing filter size $N$ of the moving average filter and corresponding statistical distributions.}
\label{fig:results_skew}
\end{figure}

A measurement series consisting of 1,000 samples was conducted using the described methodology to ensure statistical reliability. Figure~\ref{fig:results_skew} illustrates the convergence of the skew estimation for three different filter sizes, denoted by  $N$, (see Figure~\ref{fig:1_1}) as well as the statistical distribution of all tested values (see Figure~\ref{fig:1_2}).
It is visible, that the filter for $N=1024$ converges at 1.36 ppm, while skew estimation of $N=32$ jumps between 0.75 
ppm and 2 ppm. The detailed statistical characteristics of the estimated clock skew $\hat{\gamma}$ for various moving average filter sizes are summarized in Table~\ref{tab:transponiert-gerundet}. 
\begin{table}[t]
\caption{Statistical measurements moving average filter of size $N$ for estimated clock skew $\hat{\gamma}$.}
\centering
\begin{tabular}{cccccc} 
    \toprule
$\boldsymbol{N}$ & $\boldsymbol{\tilde{x}}$ & $\boldsymbol{\bar{x}}$ & $\boldsymbol{\sigma}$ & \textbf{2$\boldsymbol{\sigma}$} & \textbf{3$\boldsymbol{\sigma}$} \\
    \midrule
\textbf{1}    & 0.0000 & 1.3145 & 9.2491 & 18.4982 & 27.7472 \\
\textbf{2}    & 0.0000 & 1.3149 & 6.4737 & 12.9473 & 19.4210 \\
\textbf{4}    & 0.0000 & 1.3167 & 3.7437 & 7.4875  & 11.2312 \\
\textbf{8}    & 0.0000 & 1.3201 & 1.6869 & 3.3738  & 5.0607  \\
\textbf{16}   & 1.3750 & 1.3256 & 1.0149 & 2.0299  & 3.0448  \\
\textbf{32}   & 1.3438 & 1.3336 & 0.4972 & 0.9944  & 1.4916  \\
\textbf{64}   & 1.3594 & 1.3423 & 0.4002 & 0.8004  & 1.2006  \\
\textbf{128}  & 1.3516 & 1.3493 & 0.3483 & 0.6966  & 1.0449  \\
\textbf{256}  & 1.2789 & 1.3574 & 0.3323 & 0.6645  & 0.9968  \\
\textbf{512}  & 1.3125 & 1.3587 & 0.3290 & 0.6580  & 0.9871  \\
\textbf{1024} & 1.3085 & 1.3604 & 0.3282 & 0.6563  & 0.9845  \\
    \bottomrule
\end{tabular}
\label{tab:transponiert-gerundet}
\end{table}
The table reports the median $\tilde{x}$ , mean $\bar{x}$, standard deviation $\sigma$, and the intervals $2\sigma$ and $2\sigma$, with all values rounded to four decimal places. As observed, increasing the filter size $N$ significantly reduces the standard deviation $\sigma$, indicating that the moving average filter effectively suppresses high-frequency noise and fluctuations in the clock skew estimate. Specifically, $\sigma$ decreases from $9.2491$~ppm for $N=1$ to $0.3282$~ppm for $N=1024$. Correspondingly, the uncertainty intervals $2\sigma$ and $3\sigma$ shrink proportionally, demonstrating a substantial improvement in estimation precision.

The mean  shows a slight increasing trend with larger $N$, stabilizing around $1.36$~ppm for large filter sizes. This suggests that the moving average filter not only reduces variability but also yields a more stable estimate of the underlying clock skew. Furthermore, while the median is zero for small N (up to $N=8$), it aligns closely with the mean for larger $N$, indicating a more symmetric and less skewed distribution of the filtered estimates as the averaging window increases.

Overall, these results confirm that employing a larger moving average window leads to more reliable clock skew estimates. Hence, the maximum filter size is used for offset prediction. 

\subsection{Offset Measurements}
\label{subsec:results}
\begin{figure}[b]
\centering
 \begin{overpic}[width=0.999\columnwidth]{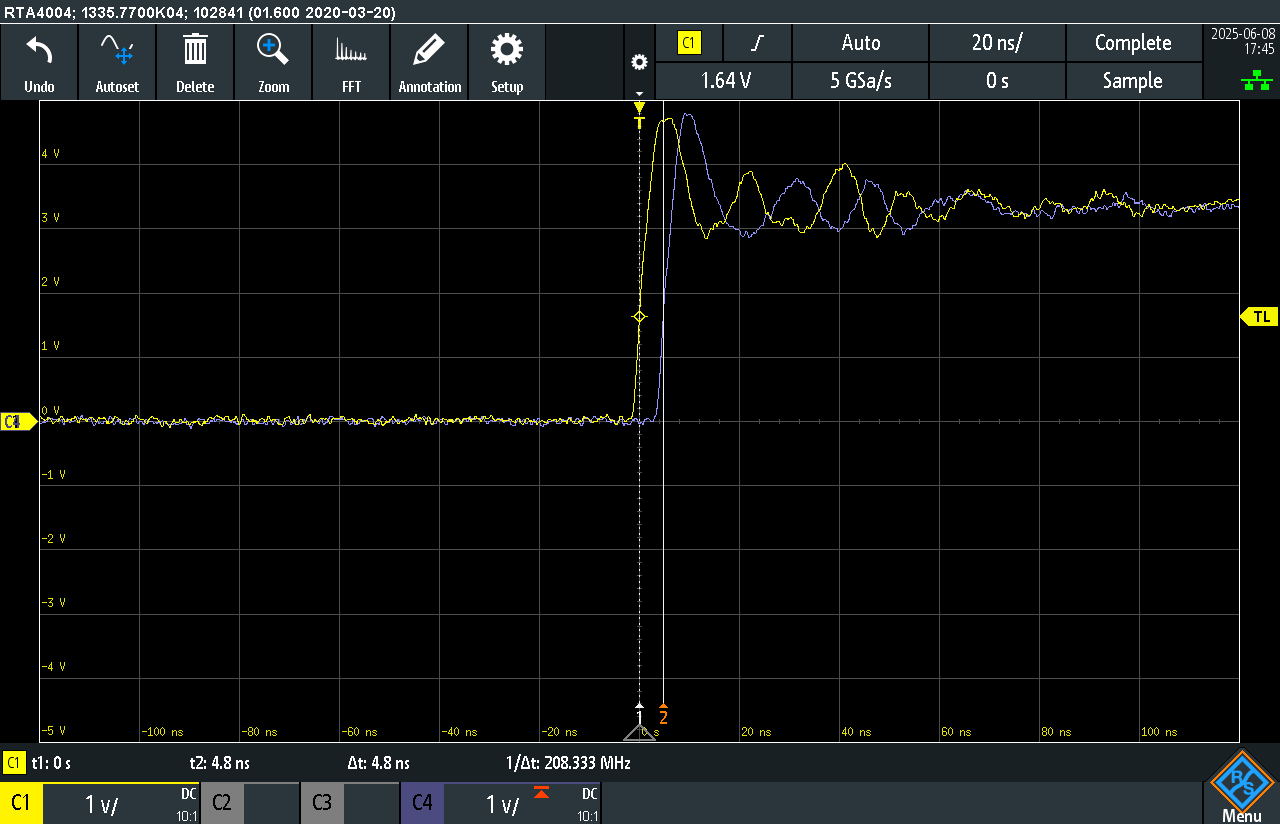}
 
 \end{overpic}
\caption{Oscilloscope screenshot showing the two rectangular validation signals captured on two channels, the trigger, and the measurement cursors.}
\label{fig:osci}
\end{figure}

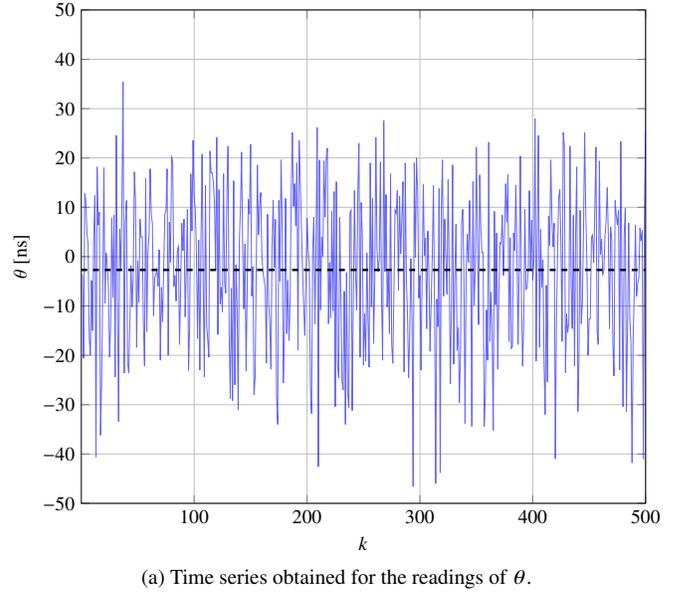
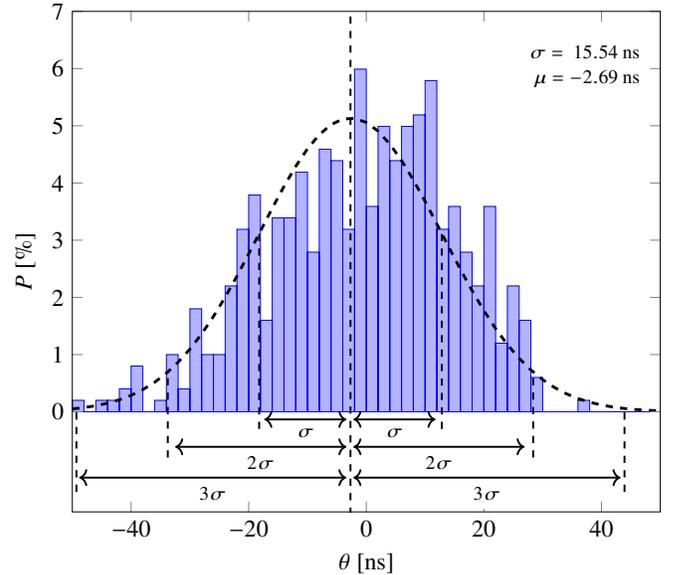
\begin{figure}[htbp]
	\centering
		\subfloat[Time series obtained for the readings of $\theta$.]{\resizebox{0.99\columnwidth}{!}{
%
%
\definecolor{mycolor1}{rgb}{0.00000,0.3,0.6}%
\begin{tikzpicture}
\begin{axis}[%
width=4in,
height=3.5in,
line width=0.05pt,
scale only axis,
xmin=0,
xmax=500,
xtick={{100},{200},{300},{400},{500}},
xlabel={$k$},
ylabel={$\theta$ [ns]},
ymin=-50,
ymax=50,
axis background/.style={fill=white},
xmajorgrids,
ymajorgrids,
yminorgrids,
]
\addplot [color=blue, opacity=0.6, forget plot]
  table[row sep=crcr]{%
1		-3.6	\\
2		-20.6	\\
3		12.8	\\
4		11.4	\\
5		4.6	\\
6		2.6	\\
7		-16.4	\\
8		-20	\\
9		-4.8	\\
10		-15	\\
11		4	\\
12		12.4	\\
13		-40.6	\\
14		18.2	\\
15		6.4	\\
16		9	\\
17		-36.2	\\
18		-25.2	\\
19		-6	\\
20		18	\\
21		-9.4	\\
22		0.8	\\
23		-1.6	\\
24		-6.2	\\
25		-20.4	\\
26		-13.8	\\
27		7.8	\\
28		-8.2	\\
29		8.4	\\
30		-24.4	\\
31		24.6	\\
32		-9	\\
33		-33.4	\\
34		5.6	\\
35		-2.6	\\
36		12	\\
37		35.4	\\
38		-23.6	\\
39		7	\\
40		11.4	\\
41		-21.6	\\
42		-23.6	\\
43		-13.2	\\
44		-1.8	\\
45		-10.2	\\
46		-3.8	\\
47		17.2	\\
48		11.4	\\
49		-18.4	\\
50		-0.8	\\
51		-9.2	\\
52		7.6	\\
53		4	\\
54		4	\\
55		-13.4	\\
56		-22.2	\\
57		15.8	\\
58		-15.2	\\
59		1	\\
60		4.2	\\
61		17.8	\\
62		8.6	\\
63		6.6	\\
64		-12	\\
65		-4.4	\\
66		-3.6	\\
67		-2.6	\\
68		-6	\\
69		1.4	\\
70		-21	\\
71		-6.2	\\
72		-13.2	\\
74		8.6	\\
75		9.2	\\
76		17.8	\\
77		-20	\\
78		7	\\
79		-1.2	\\
80		20.4	\\
81		19	\\
82		-6	\\
83		-3.8	\\
84		-9	\\
85		1	\\
86		2	\\
87		-17.8	\\
88		-6.8	\\
89		6.8	\\
90		1.2	\\
91		7.6	\\
92		-12.2	\\
93		-2.4	\\
94		9.6	\\
95		-23.2	\\
96		-14	\\
97		16.8	\\
98		5.2	\\
99		23.6	\\
100		12.2	\\
101		9.6	\\
102		0.6	\\
103		-16.6	\\
104		3.4	\\
105		-23	\\
106		2.6	\\
107		20.8	\\
108		-5.6	\\
109		-24.4	\\
110		14.4	\\
111		-13.8	\\
112		-16.8	\\
113		-20.4	\\
114		21.4	\\
115		17	\\
116		17	\\
117		15	\\
118		10.8	\\
119		-2.8	\\
120		24.2	\\
121		-15.8	\\
122		13.6	\\
123		-9.6	\\
124		-6.2	\\
125		-11.2	\\
126		16.8	\\
127		-5.4	\\
128		-7.4	\\
129		13.2	\\
130		22.4	\\
131		-16	\\
132		-28.8	\\
133		-6.4	\\
134		-29.2	\\
135		15.4	\\
136		-26	\\
137		-1.6	\\
138		-10.2	\\
139		-31	\\
140		3	\\
141		9.8	\\
142		21.2	\\
143		1.4	\\
144		-10.6	\\
145		2.6	\\
146		-23.2	\\
147		-6.4	\\
148		11.6	\\
149		8.6	\\
150		22.8	\\
151		-8	\\
152		-2	\\
153		-28	\\
154		-24.6	\\
155		-2.8	\\
156		18.6	\\
157		4.4	\\
158		10.6	\\
159		13	\\
160		0.2	\\
161		-3.2	\\
162		-4.8	\\
163		-6.4	\\
164		-21.6	\\
165		8.6	\\
166		-9.4	\\
167		-13.4	\\
168		-22.8	\\
169		1.8	\\
170		11.4	\\
171		-2.2	\\
172		7.6	\\
173		-28.2	\\
174		-34	\\
175		11.4	\\
176		-5	\\
177		19.6	\\
178		5.8	\\
179		-25.6	\\
180		-13.4	\\
181		11.8	\\
182		-15.6	\\
183		-4.8	\\
184		-17	\\
185		-9	\\
186		13.2	\\
187		25.2	\\
188		10.2	\\
189		14.8	\\
190		6.2	\\
191		19	\\
192		-19	\\
193		23.6	\\
194		15.8	\\
195		6	\\
196		-2.6	\\
197		-9.8	\\
198		-16	\\
199		-4.6	\\
200		7.8	\\
201		1.4	\\
202		-1.8	\\
203		-26.2	\\
204		-31.8	\\
205		1	\\
206		8	\\
207		-13.6	\\
208		4.2	\\
209		26.2	\\
210		-42.6	\\
211		19.6	\\
212		-10.8	\\
213		4	\\
214		0	\\
215		19.4	\\
216		8	\\
217		19.8	\\
218		22	\\
219		-13.6	\\
220		-1.2	\\
221		-12.4	\\
222		9.2	\\
223		-11	\\
224		13.2	\\
225		-30.4	\\
226		15.2	\\
227		-21.8	\\
228		-24.6	\\
229		8	\\
230		-8.6	\\
231		-25.2	\\
232		-27	\\
233		-7	\\
234		-34	\\
235		-3.2	\\
236		-27.8	\\
237		-30.6	\\
238		8.8	\\
239		9.4	\\
240		-31.2	\\
241		-9.8	\\
242		13.4	\\
243		-3.6	\\
244		-10.8	\\
245		13	\\
246		23	\\
247		-20.4	\\
248		11	\\
249		4.4	\\
250		-22	\\
251		-11.6	\\
252		-21.2	\\
253		0.8	\\
254		-12.4	\\
255		-22.4	\\
256		19	\\
257		-10.2	\\
258		11.8	\\
259		-17.6	\\
260		-5	\\
261		24.2	\\
262		1.6	\\
263		-3	\\
264		-0.6	\\
265		19.2	\\
266		-8.2	\\
267		-21	\\
268		27.6	\\
269		-3	\\
270		12.6	\\
271		-10.2	\\
272		-3	\\
273		-15.6	\\
274		16.8	\\
275		-17.8	\\
276		-21.6	\\
277		-7	\\
278		9.4	\\
279		8.8	\\
280		13.4	\\
281		-6.8	\\
282		12.2	\\
283		0.6	\\
284		7.8	\\
285		15.2	\\
286		-23.4	\\
287		19	\\
288		-4.6	\\
289		-11.6	\\
290		-24.6	\\
291		-15.6	\\
292		-0.4	\\
293		-12.4	\\
294		-46.6	\\
295		19	\\
296		-0.2	\\
297		20	\\
298		13	\\
299		-8.4	\\
300		-3.8	\\
301		-23	\\
302		-4	\\
303		14.6	\\
304		0.2	\\
305		3.8	\\
306		-14.6	\\
307		5.2	\\
308		-3.2	\\
309		-19.2	\\
310		-9	\\
311		0.2	\\
312		14.4	\\
313		-10.4	\\
314		-46	\\
315		-8.6	\\
316		-14.2	\\
317		13.8	\\
318		-43.8	\\
319		10	\\
320		19.6	\\
321		-8	\\
322		7.6	\\
323		-0.6	\\
324		-7.6	\\
325		15.6	\\
326		-2.8	\\
327		-25.8	\\
328		-9.6	\\
329		18.6	\\
330		-21.8	\\
331		-3.4	\\
332		6.6	\\
333		-19	\\
334		-14.4	\\
335		-29.6	\\
336		-19.2	\\
337		7	\\
338		9.6	\\
339		-3	\\
340		-33.8	\\
341		-3	\\
342		4.6	\\
343		-1.2	\\
344		11.4	\\
345		1.2	\\
346		-34.4	\\
347		15.2	\\
348		-9.8	\\
349		-2.8	\\
350		22.2	\\
351		3.2	\\
352		-13.8	\\
353		16.6	\\
354		5.8	\\
355		0.8	\\
356		0.8	\\
357		-34.4	\\
358		-27.8	\\
359		-10.8	\\
360		-21	\\
361		23.2	\\
362		-17.2	\\
363		-7.2	\\
364		-15.6	\\
365		-35.2	\\
366		9.2	\\
367		-9.8	\\
368		4.8	\\
369		-22.8	\\
370		-7.2	\\
371		2.8	\\
372		2.6	\\
373		8.4	\\
374		-3.8	\\
375		4.6	\\
376		14.8	\\
377		7	\\
378		16.6	\\
379		-17.2	\\
380		3.8	\\
381		-18.6	\\
382		-18.2	\\
383		4.8	\\
384		1.2	\\
385		14.4	\\
386		1.2	\\
387		-17.8	\\
388		-1.8	\\
389		20.4	\\
390		-8.4	\\
391		-22.2	\\
392		-14.2	\\
393		9.6	\\
394		-6.6	\\
395		-6.4	\\
396		-2.4	\\
397		13.4	\\
398		-2.8	\\
399		7.4	\\
400		-4.8	\\
401		-0.8	\\
402		28	\\
403		-21.2	\\
404		5.4	\\
405		24.6	\\
406		-18.4	\\
407		5.6	\\
408		-6.2	\\
409		-0.6	\\
410		-21.4	\\
411		-32	\\
412		-5.8	\\
413		-25.4	\\
414		8.2	\\
415		2.2	\\
416		19.6	\\
417		7.4	\\
418		2	\\
419		6.8	\\
420		-41	\\
421		-13.4	\\
422		1.2	\\
423		12	\\
424		13.6	\\
425		2.4	\\
426		-17.2	\\
427		25.2	\\
428		22.4	\\
429		-23.6	\\
430		-11.6	\\
431		-15.4	\\
432		-13.4	\\
433		22.4	\\
434		-4.6	\\
435		-14.6	\\
436		8.4	\\
437		-22.2	\\
438		4.8	\\
439		18.2	\\
440		-31.4	\\
441		-20.4	\\
442		9.4	\\
443		-13.6	\\
444		-9.2	\\
445		6	\\
446		25.2	\\
447		8.6	\\
448		-1.2	\\
449		-20	\\
450		-12.6	\\
451		-12.6	\\
452		4	\\
453		4.4	\\
454		-1.2	\\
455		9.6	\\
456		22.2	\\
457		-14.8	\\
458		-17.8	\\
459		19.4	\\
460		0.4	\\
461		3.6	\\
462		-3.6	\\
463		-2.8	\\
464		4	\\
465		7.4	\\
466		14	\\
467		4	\\
468		-14.6	\\
469		6	\\
470		3.4	\\
471		-17.8	\\
472		-24.2	\\
473		11.4	\\
474		8.4	\\
475		6.4	\\
476		9.4	\\
477		-23	\\
478		23.4	\\
479		-7	\\
480		-30.4	\\
481		4.4	\\
482		9.8	\\
483		-31.4	\\
484		-14.2	\\
485		6.8	\\
486		-12.6	\\
487		-17.4	\\
488		-41.8	\\
489		-16	\\
490		-7	\\
491		6.4	\\
492		-8	\\
493		-5.4	\\
494		-4.4	\\
495		5.8	\\
496		3.2	\\
497		5.4	\\
498		-41	\\
499		10	\\
500		25.4	\\
};
 \addplot [color=black,dashed, style={very thick}, forget plot]
  table[row sep=crcr]{%
  0 -2.69  \\
  500   -2.69 \\
 };
\end{axis}
\end{tikzpicture}%
}\label{fig:2_1}}
        
		 \subfloat[Histogram of the probability density function obtained for the readings of $\theta$.]{\resizebox{0.99\columnwidth}{!}{\definecolor{mycolor1}{rgb}{0.00000,0.3,0.6}
\definecolor{mycolor3}{rgb}{0.01,0.79,0.395}

\begin{tikzpicture}
\begin{axis}[ 
width=4in,
height=3.5in,
ymin=-1.75, ymax=7,
xmin=-50, xmax=50,
area style,
ytick={0,1,2,3,4,5,6,7,8,9,10},
ylabel={$P$ [$\mathrm{\%}$]},
xlabel={$\theta$ [ns]},
    ]
\addplot+[ybar interval,mark=no] plot coordinates { 
(	-50	,	0.2	)
(	-48	,	0.00E+00	)
(	-46	,	0.2	)
(	-44	,	0.2	)
(	-42	,	0.4	)
(	-40	,	0.8	)
(	-38	,	0	)
(	-36	,	0.2	)
(	-34	,	1	)
(	-32	,	0.4	)
(	-30	,	1.8	)
(	-28	,	1	)
(	-26	,	1	)
(	-24	,	2.2	)
(	-22	,	3.19	)
(	-20	,	3.79	)
(	-18	,	1.6	)
(	-16	,	3.39	)
(	-14	,	3.39	)
(	-12	,	4.19	)
(	-10	,	2.79	)
(	-8	,	4.59	)
(	-6	,	4.39	)
(	-4	,	3.19	)
(	-2	,	5.99	)
(	0	,	3.59	)
(	2	,	4.99	)
(	4	,	4.39	)
(	6	,	4.99	)
(	8	,	5.19	)
(	10	,	5.79	)
(	12	,	3.19	)
(	14	,	3.59	)
(	16	,	2.79	)
(	18	,	2.2	)
(	20	,	3.59	)
(	22	,	1.2	)
(	24	,	2.2	)
(	26	,	1.6	)
(	28	,	0.6	)
(	30	,	0	)
(	32	,	0	)
(	34	,	0	)
(	36	,	0.2	)
(	38	,	0	)
(	40	,	0	)
(	42	,	0	)
(	44	,	0	)
(	46	,	0	)
(	48	,	0	)
(	50	,	0	)
};

\addplot [color=black, style={very thick}, dashed,forget plot]
  table[row sep=crcr]{%
-50		0.0497713	\\
-48		0.073049226	\\
-46		0.105452329	\\
-44		0.149727185	\\
-42		0.20909761	\\
-40		0.287211221	\\
-38		0.388023173	\\
-36		0.515605861	\\
-34		0.673879049	\\
-32		0.866263443	\\
-30		1.0952719	\\
-28		1.362064984	\\
-26		1.666010123	\\
-24		2.004293733	\\
-22		2.371641415	\\
-20		2.760200358	\\
-18		3.159629171	\\
-16		3.55742331	\\
-14		3.939480027	\\
-12		4.290878174	\\
-10		4.596818859	\\
-8		4.843647409	\\
-6		5.019859675	\\
-4		5.116989875	\\
-2		5.130284659	\\
0		5.059088692	\\
2		4.906898202	\\
4		4.681076386	\\
6		4.392262957	\\
8		4.053543725	\\
10		3.679470423	\\
12		3.285032688	\\
14		2.884682363	\\
16		2.491496193	\\
18		2.116539517	\\
20		1.768464962	\\
22		1.453350829	\\
24		1.174757944	\\
26		0.93396423	\\
28		0.730324754	\\
30		0.56170168	\\
32		0.424912346	\\
34		0.316152705	\\
36		0.231365382	\\
38		0.166534332	\\
40		0.11789981	\\
42		0.082096824	\\
44		0.056226822	\\
46		0.037876046	\\
48		0.025095142	\\
50		0.016353797	\\
};
\node[anchor=north east] at (rel axis cs:0.98,0.95){\footnotesize{$\sigma=~15.54~\mathrm{ns}$}};
\node[anchor=north east] at (rel axis cs:0.98,0.90){\footnotesize{$\mu=-2.69~\mathrm{ns} $}};
 \addplot [color=black,dashed, style={thick}, forget plot]
  table[row sep=crcr]{%
  -2.69   -1.8 \\
  -2.69    8\\
 };
  \addplot [color=black,dashed, style={thick}, forget plot]
  table[row sep=crcr]{%
  -49.30   -1.4 \\
  -49.30    0\\
 };
  \addplot [color=black,dashed, style={thick}, forget plot]
  table[row sep=crcr]{%
  -33.76  -0.8 \\
  -33.76   0.7
  \\
 };
  \addplot [color=black,dashed, style={thick}, forget plot]
  table[row sep=crcr]{%
  -18.22   -0.3 \\
  -18.22    3.11\\
 };
  \addplot [color=black,dashed, style={thick}, forget plot]
  table[row sep=crcr]{%
  12.85   -0.3 \\
  12.85    3.11\\
 };
  \addplot [color=black,dashed, style={thick}, forget plot]
  table[row sep=crcr]{%
  28.39   -0.8 \\
  28.39    0.7\\
 };
  \addplot [color=black,dashed, style={thick}, forget plot]
  table[row sep=crcr]{%
  43.92   -1.4 \\
  43.92    0\\
 };
\end{axis}
    \draw[<->, thick] (0.1,0.5) -- (4,0.5) node[midway, below] {\footnotesize$3\sigma$};
\draw[<->, thick] (1.5,0.95) -- (4,0.95) node[midway,below] {\footnotesize2$\sigma$};
\draw[<->, thick] (2.8,1.35) -- (4,1.35) node[midway, below] {\footnotesize$\sigma$};

\draw[<->, thick] (4.1,1.35) -- (5.3,1.35) node[midway, below] {\footnotesize$\sigma$};
\draw[<->, thick] (4.1,0.95) -- (6.6,0.95) node[midway, below] {\footnotesize2$\sigma$};
\draw[<->, thick] (4.1,0.5) -- (8,0.5) node[midway, below,fill=white] {\footnotesize3$\sigma$};
\end{tikzpicture}}\label{fig:2_2}}
\caption{Results of the measurements of the offset between the two rectangular signals.}
\label{fig:results_offsets}
\end{figure}

A screenshot of the oscilloscope displaying the two rectangular signals that are used for evaluation on Channel~1~(C1) and Channel~4~(C4) is shown in Figure~\ref{fig:osci}, 
with the trigger level set to half of the maximum signal voltage. In this screenshot, the measurement cursors indicate that the temporal offset between the two signals is approximately \SI{4.8}{\nano\second}. To derive statistical probabilities, a series of 500 measurements was carried out and analyzed. The number of measurements was limited to 500 because a preliminary analysis showed that increasing the sample size beyond this point did not significantly improve the statistical power or reduce variance further. Collecting more data would have required disproportionately higher experimental effort and processing time without providing additional insights. The results are depicted in Figure~\ref{fig:results_offsets}, whereas Figure~\ref{fig:2_1} shows the time series and Figure~\ref{fig:2_2} both histogram and probability density function of the offset that has been measured by the oscilloscope.
Additionally, Table~\ref{tab:probabilities} presents the precision in terms of the measured clock offset its associated standard deviations. 
\begin{table}[htbp]
    \centering
    \caption{Measured clock precision for different standard deviations.}
    \begin{tabular}{c   c   c   c}
    \toprule
   & $\sigma$ & $2\sigma$  & $3\sigma$ \\
     \midrule
  $E(\theta=\upmu)$ [ns]   & -2.69$\pm$15.54  & -2.69$\pm$31.07 & -2.69$\pm$46.61 \\
   $P [\mathrm{\%}]$ & 68.27& 95.45& 99.73\\
      \bottomrule 
    \end{tabular}
    \label{tab:probabilities}
\end{table}
The mean clock offset is approximately \SI{2.69}{\nano\second}, indicating a slight systematic bias, while the spread of the measurements increases with higher standard deviation intervals. Specifically, for 1$\sigma$ the offset is within $\pm$\SI{15.54}{\nano\second} with a probability $P=68.27$~\%. For 2$\sigma$, the interval widens to $\pm$\SI{31.07}{\nano\second} with a probability of 95.45~\%, and for 3$\sigma$, the maximum observed spread is $\pm$\SI{46.61}{\nano\second} with a probability of 99.73~\%. 
These results confirm that the proposed synchronization scheme achieves high precision, with the vast majority of offset estimates confined to a narrow interval of less than \SI{50}{\nano\second} even under very conservative confidence levels. The small mean offset further indicates minimal systematic error in the clock alignment.

\section{Conclusion}%
\label{sec:Conclusion}
This paper tackled the challenge of achieving sub-microsecond time synchronization in wireless systems, focusing on industrial applications and \acrshort{5g} networks. We proposed and enhanced a robust protocol capable of synchronizing \glspl{ue}, \gls{gnb}, and \gls{5gc}, making it suitable for hybrid network deployments. The solution was implemented on a hardware testbed using \gls{oai} and  \glspl{sdr}, with performance evaluated through moving average filters of varying sizes. Results showed that a filter size of 1024 yields the highest precision by effectively reducing high-frequency noise and stabilizing clock skew estimates.

Experimental validation in a controlled laboratory environment demonstrated synchronization precision within $\pm$\SI{50}{\nano\second}, which is an order of magnitude better than the one microsecond target, highlighting the approach’s feasibility for high-precision industrial \acrshort{5g} and 6G networks. While real-world factors such as propagation delay may introduce additional challenges, adding \gls{ta} information can compensate for these effects. Overall, the achieved precision provides a strong margin to meet the stringent timing requirements of time-sensitive industrial applications, confirming the effectiveness of the proposed synchronization method for practical deployment.


\balance
\printbibliography

%
%
\end{document}